# Tunnel magnetoresistance angular and bias dependence enabling tuneable wireless communication

Ewa Kowalska[1,2]*, Akio Fukushima[3], Volker Sluka[1], Ciarán Fowley[1], Attila Kákay[1], Yuriy Aleksandrov[1,2], Jürgen Lindner[1], Jürgen Fassbender[1,2], Shinji Yuasa[3], Alina M. Deac[1]**

[1] Helmholtz-Zentrum Dresden-Rossendorf, Institute of Ion Beam Physics and Materials Research, Dresden, Germany

[2] Institute of Solid State Physics, TU Dresden, Germany

[3] National Institute of Advanced Industrial Science and Technology, Spintronics Research Center, Tsukuba, Japan

*e.kowalska@hzdr.de
**a.deac@hzdr.de

**Spin-transfer torques (STTs) can be exploited in order to manipulate the magnetic moments of nanomagnets, thus allowing for new consumer-oriented devices to be designed. Of particular interest here are tuneable radio-frequency (RF) oscillators for wireless communication. Currently, the structure that maximizes the output power is an Fe/MgO/Fe-type magnetic tunnel junction (MTJ) with a fixed layer magnetized in the plane of the layers and a free layer magnetized perpendicular to the plane. This structure allows for most of the tunnel magnetoresistance (TMR) to be converted into output power. Here, we experimentally and theoretically demonstrate that the main mechanism sustaining steady-state precession in such structures is the angular dependence of the magnetoresistance. The TMR of such devices is known to exhibit a broken-linear dependence versus the applied bias. Our results show that the TMR bias dependence effectively quenches spin-transfer-driven precession and introduces a non-monotonic frequency dependence at high applied currents. Thus we expect the bias dependence of the TMR to have an even more dramatic effect in MTJs with Mn-Ga-based free layers, which could be used to design wireless oscillators extending towards the 'THz gap', but have been experimentally shown to exhibit a non-trivial TMR bias dependence.**

## Introduction

While initial spin-torque nano-oscillators (STNOs) studies focused on devices with fully in-plane (IP) magnetized magnetic layers[1,2], hybrid device geometries combining an IP reference layer and an out-of-plane (OOP) magnetized free layer are now the system of choice in view of potential applications[3-7]. Such a system is sketched in Fig. 1 (a), with the free layer having an easy axis along the perpendicular to plane ($z$) direction, and the reference layer magnetized along an in-plane direction, defined here as the $x$-axis. This configuration maximizes the output power, reduces the critical current[8], and can allow for steady-state precession to be excited regardless of applied current or magnetic field history[3,9,10]. State-of-the-art devices, exploiting Fe/MgO/Fe-based MTJs[11,12], can exhibit output powers orders of magnitude higher (as high as µW[13,14]) than their fully metallic giant magnetoresistance (GMR) predecessors (limited to a few nW). Simultaneously, these nano-oscillators have a lateral size about 50 times smaller than devices presently used in mobile telecommunication[15]. TMR devices further benefit from low operational current densities of the order of 1 MA/cm$^2$ [4,9,16], i.e. one order of magnitude lower as compared to the case of metallic spin-valves[2,3]. A device with the same hybrid geometry was recently integrated into phase-locked-loops exhibiting extremely narrow linewidth (less than 1 Hz), which make them suitable for wireless communication applications.[17]

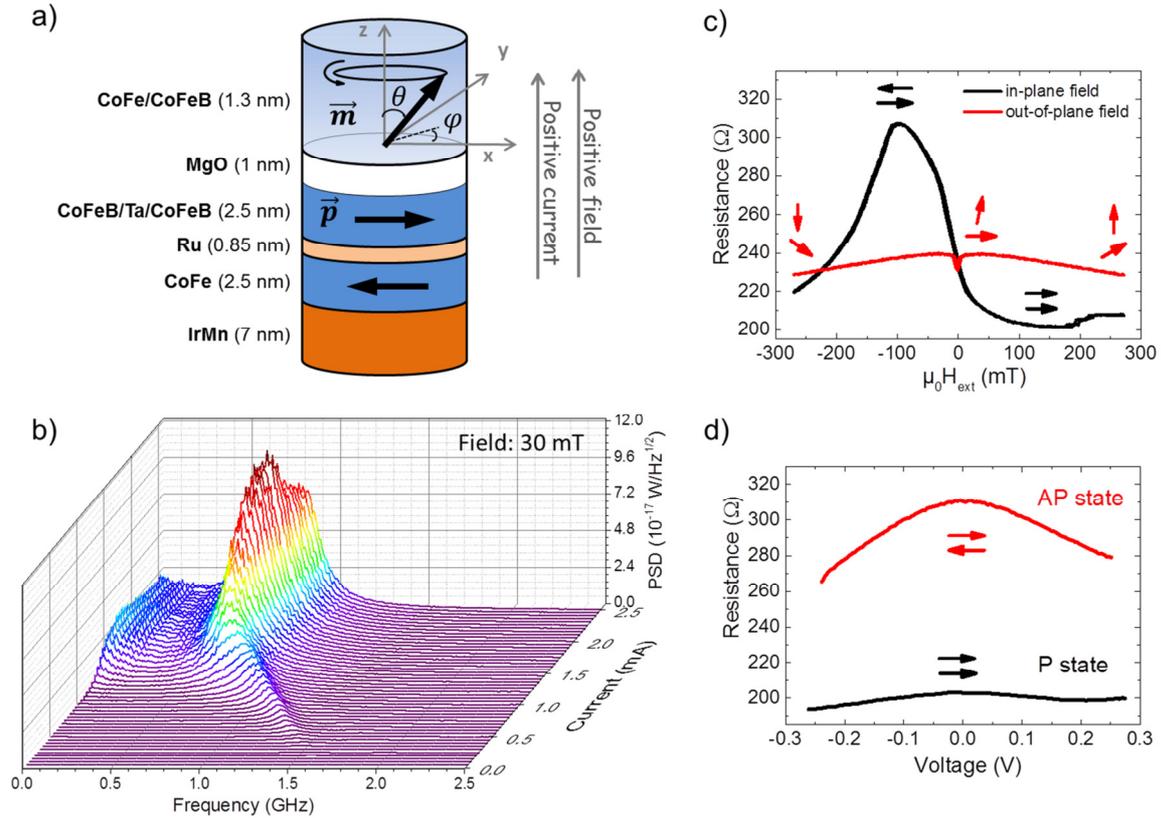

**Figure 1. Scheme of the STNO sample, and its static, dynamic and frequency characteristics.** (**a**), STNO sample with marked magnetization orientations of the free layer and reference layer (**m** and **p**, respectively), directions of positive applied current and magnetic field, and the coordinate system used in the calculations. The free layer magnetization **m** is depicted precessing at an angle $\theta$ around the $z$-axis, under the assumption that no other anisotropies are present in the system. In the experiment, an applied DC current, via spin-transfer torque, induces precession of the free layer magnetization, which leads to a time-varying voltage, detected via a spectrum analyzer. (**b**), Frequency spectra versus DC current at a field of 30 mT showing a non-monotonic frequency variation and a decrease of the output power at large currents (above 2.2 mA). (**c**), Magnetoresistance curves measured with in-plane and an out-of-plane magnetic fields ($\mu_0 H_{ext}$), exhibit the characteristic behaviour of MTJs with a geometry as shown in (a)[5]. The magnetization directions of the free and the reference layers are depicted by the upper and lower arrows, respectively. (**d**), Resistance versus bias voltage, showing a linear decrease in the resistance for the antiparallel (AP) state and an approximately constant resistance for the parallel (P) state.

For both GMR and TMR devices, the STT-driven magnetization dynamics can be described by the Landau-Lifshitz-Gilbert-Slonczewski (LLGS) equation[18]:

$$\frac{d\boldsymbol{m}}{dt} = -\gamma\mu_0(\boldsymbol{m}\times\boldsymbol{H_{eff}}) + \alpha\left(\boldsymbol{m}\times\frac{d\boldsymbol{m}}{dt}\right) + \gamma T_\parallel[\boldsymbol{m}\times(\boldsymbol{m}\times\boldsymbol{p})] + \gamma T_\perp(\boldsymbol{m}\times\boldsymbol{p}). \qquad (1)$$

Here, $\gamma$ is the gyromagnetic ratio, **m** is the unit vector along the magnetization direction, $\boldsymbol{H_{eff}}$ is the effective magnetic field, $\alpha$ is the Gilbert damping constant, **p** is the unit vector defining the direction of the spin-polarization of the current, $T_\parallel$ is the in-plane (Slonczewski) spin-transfer torque, and $T_\perp$ is the out-of-plane (field-like) spin-transfer torque.

In GMR devices, $T_\parallel^{GMR} = \frac{\hbar}{2e}\frac{1}{M_s\Omega}g(\theta,\varphi,\lambda)I_{DC}$ (where $M_s$ is the saturation magnetization, $\Omega$ is the magnetic volume, $I_{DC}$ is the applied constant current and $g(\theta,\varphi,\lambda)$ is the asymmetry factor), while $T_\perp^{GMR}$ is negligible.[19] The underlying mechanism responsible for sustaining dynamics in this hybrid geometry is the spin-transfer torque angular asymmetry, expressed by the asymmetry parameter $\lambda$[19]. The angular asymmetry of $T_\parallel^{GMR}$ leads to a *net* Slonczewski torque when integrating over a full precession cycle (as defined by the effective field) for electrons flowing from the free to the reference layer, and therefore allows for compensating the damping torque at sufficiently large currents[3,20]. MTJs have been experimentally shown to exhibit similar dynamics, in spite of the fact that $T_\parallel^{MTJ} = a_\parallel V$, where $V$ is the applied bias voltage and $a_\parallel$ is a constant known as the in-plane torkance and, therefore, the Slonczewski torque cancels on one precession cycle and cannot counteract the damping. Note that in MTJs the perpendicular STT term, $T_\perp^{MTJ} = a_\perp V^2$, is finite (here, $a_\perp$ is the field-like torkance). [21-23]

In this article, we experimentally and theoretically investigate spin-transfer-driven dynamics in MgO-based MTJs considering the voltage bias dependence of the magnetoresistance and the spin-transfer torques. Experimentally, we observe an unusual, but reproducible curvature of the critical lines in the current-field phase diagram enclosing the region of steady-state dynamics which, to the authors knowledge, has never been reported in similar metallic- or MTJ-based devices. Theoretically, we incorporate the angular dependence of the TMR[24-26] and bias dependence[5,21,27] into spin-transfer torque terms, $T_\parallel^{MTJ}$ and $T_\perp^{MTJ}$, and then solve equation (1). We find that the angular dependence of the resistance in MTJs, alone, introduces an asymmetry in $T_\parallel^{MTJ}$ and gives rise to steady-state precession. Moreover, including the bias dependence of TMR correctly reproduces the curvature of the regions of steady-state precession in the experimental phase diagram. Furthermore, we also examine the effect of the bias dependence of TMR on $T_\parallel^{MTJ}$ and show that it gradually suppresses the induced asymmetry, which ultimately leads to the quenching of dynamics at high bias currents. The analytical formalism presented here allows for the estimation of achievable and realistic device parameter values for driving spin-torque dynamics in MTJ stacks with efficiencies in excess of what can be achieved in GMR devices. Therefore, the TMR ratio, as well as its bias dependence, which are generally not taken into account, are both equally crucial factors governing the performance of MTJ-based STNOs.

**Results and discussion**

The experimental geometry and measurement scheme of our STNO devices with marked directions of applied DC current, $I_{DC}$, and magnetic field, $\mu_0 H_{ext}$, is presented in Fig. 1 (a). Multilayer films of the following composition: buffer layers / IrMn 7 / $Co_{70}Fe_{30}$ 2.5 / Ru 0.85 / $Co_{20}Fe_{60}B_{20}$ 1.2 / Ta 0.2 / $Co_{20}Fe_{60}B_{20}$ 1.2 / $Co_{30}Fe_{70}$ 0.4 / MgO 1 / $Co_{30}Fe_{70}$ 0.2 / $Co_{20}Fe_{60}B_{20}$ 1.1 / capping layers (thicknesses in nm) were grown on $Si/SiO_2$ substrates and patterned into nano-pillar devices with the same magnetic volume, but with varying cross sections, by a combination of ultraviolet and electron beam lithography. While all devices show similar trends, here, we present experimental data from a representative device patterned to a (250 x 50) nm ellipse (Fig. 1 (b,c,d) and 2 (a,b)).

Fig. 1 (b) shows measured STNO frequency spectra versus applied DC current at a 30 mT applied field. We observed a non-monotonic frequency variation and a decrease of the output power occurring for currents above 2.2 mA. The MTJ resistance as a function of $\mu_0 H_{ext}$, applied both in the plane of the layers (along the $x$-axis) and along the normal, is shown in Fig. 1 (c). The curves confirm that the effective magnetic anisotropy of the free layer, $\mu_0(H_k - M_s)$, is greater than zero and thus its magnetic easy axis lies out of the film plane[5]. The bias dependence of the resistance, Fig. 1 (d), shows a decrease in the resistance for the AP state, with a slope $\frac{\partial R_{AP}}{\partial V}$ = 105 $\frac{\Omega}{V}$ (for positive voltage driving dynamics), and approximately constant resistance for the P state. For a given range of $I_{DC}$ and $\mu_0 H_{ext}$ applied along the film normal, precession of the free layer magnetization is excited. This yields an oscillatory resistance and, hence, voltage which can be directly detected by means of a spectrum analyzer. The output power of the STNO is then obtained by integration of the fitted peak area (for details, see Supplementary Information: Part 1).

The phase diagram of the observed dynamics as a function of $\mu_0 H_{ext}$ (applied along the $z$ direction) and $I_{DC}$ is shown in Fig. 2 (a). We find that, similar to metallic devices[3], spin-torque driven steady-state dynamics can be obtained only for electrons flowing from the free to the reference layer (defined here as positive current), also in agreement with previous reports on MgO-based MTJs[4,16]. The colour code represents the integrated output power, reaching a maximum of 55 nW (obtained for $\mu_0 H_{ext}$ = 7 mT and $I_{DC}$ = 1.9 mA). A clear curvature can be observed at the boundary of the regions of high power steady-state dynamics, denoted with the dashed white line. At low currents, this curvature is quasi-parabolic with increasing external field. Above 2.2 mA the dynamics are gradually quenched, and by extrapolation are assumed to have totally decayed above 2.7 mA. The maximum applied current of 2.5 mA refers to a voltage of 0.775 V (slightly lower than the breakdown voltage of the device, approximately 0.8 V in the AP state).

Magnetoresistance curves are recorded simultaneously with the microwave emission. This allows us to monitor the time-averaged static resistance which is proportional to the projection of the free layer magnetization on the reference layer, as shown in Fig. 2 (b). We define the change in resistance $\Delta R'$ with respect to the resistance at the same field when using a small probe current which does not stimulate strong dynamics, $\Delta R'|_{\mu_0 H_{ext}} = R(I_{DC}) - R(0.5\ mA)$. In this way, the value of $\Delta R'$ is directly proportional to the average $m_x$ component of the magnetization (for details see Supplementary Information: Part 2). As shown in Fig. 2 (b), for field values close to zero, the spin-torque stabilizes the static in-plane AP state[6]. Note also that, according to panels (a) and (b) in Fig. 2, magnetization dynamics are only observable in the presence of finite external fields.

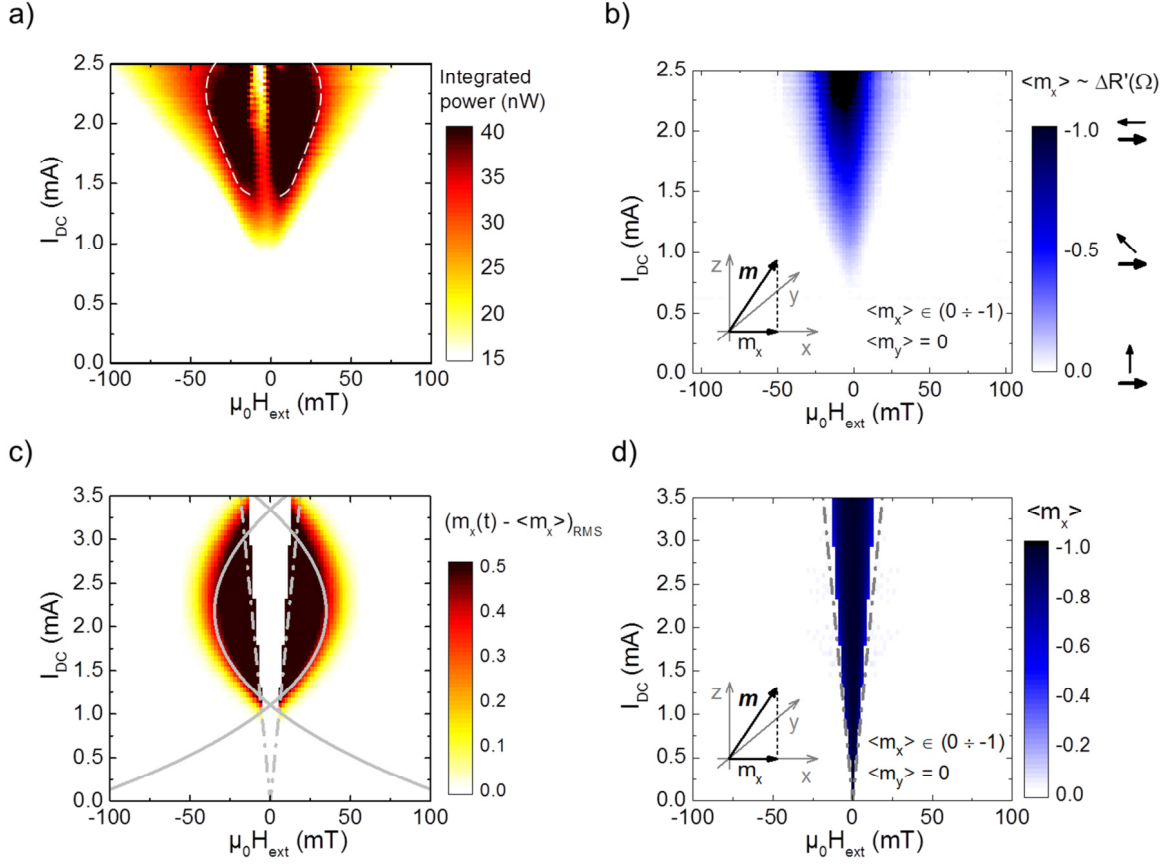

**Figure 2. Experimentally (a,b) and numerically (c,d) obtained dynamic and static characteristics of a typical nano-oscillator sample.** (**a**), Measured output power of the STNO as a function of DC current, $I_{DC}$, and $\mu_0 H_{ext}$. The white dashed line is a guide to the eye marking the strong curvature around the high power region which appears for currents above 2.2 mA. (**b**), Time-averaged projection of the free layer magnetization along the reference layer magnetization ($m_x$), as function of $I_{DC}$ and $\mu_0 H_{ext}$, determined from static resistance measurements. The static in-plane AP state (black region) is stabilized for currents above 2.0 mA. Data in (a) is recorded simultaneously with (b). (**c**), Computed dynamics intensity of the STNO along $x$-axis as a function of $I_{DC}$ and $\mu_0 H_{ext}$ taking into account a linear dependence of TMR on the applied bias ($\frac{\partial R_{AP}}{\partial V} = 105 \frac{\Omega}{V}$). Solid lines show analytically determined critical currents for dynamics and dashed lines show the boundaries marking the stability of the static in-plane AP state. The linear bias dependence of TMR, defined by $\frac{\partial R_{AP}}{\partial V}$, is responsible for the curvature of the critical lines. A gradual quenching of dynamics above 2.2 mA is observed with complete suppression of spin-transfer induced dynamics above 3.5 mA. (**d**), Numerically obtained average $m_x$ component as a function of $I_{DC}$ and $\mu_0 H_{ext}$, determining the stability region of the static in-plane AP state. Dashed lines show analytically determined boundaries of the static in-plane AP state.

We describe the system analytically using equation (1) and introduce the effective field as $H_{eff} = H_{ext} + H_{k_\perp} m_z$, where $H_{ext}$ is the applied external field, $H_{k_\perp}$ is the out-of-plane effective anisotropy, and $m_z$ is the projection of the unit vector **m** along $z$.

As experiments are conducted at constant current, the appropriate voltage $V$ (which is then introduced into $T_\parallel^{MTJ}$) is related to $I_{DC}$ via the following:

$$V(\theta,\varphi) = I_{DC} R(\theta,\varphi) = I_{DC} \frac{R_P + \frac{\Delta R_0}{2}(1-\sin\theta\cos\varphi)}{1+\frac{1}{2}|I_{DC}|\frac{\partial R_{AP}}{\partial V}(1-\sin\theta\cos\varphi)}. \quad (2)$$

Here, $R(\theta,\varphi)$ is the instantaneous resistance, $R_P$ is the P state resistance, $\Delta R_0$ is the resistance difference between P and AP state close to zero bias, $\theta$ and $\varphi$ are the angles of spherical coordinate system which define the position of the magnetization vector and, thus, $\sin\theta\cos\varphi$ is the projection of the free layer magnetization vector on the direction of the reference layer (i. e., $\mathbf{m}\cdot\mathbf{p}$), and $\frac{\partial R_{AP}}{\partial V}$ is the slope of the bias dependence of the AP state resistance.

The following assumptions are made: a linear and quadratic voltage dependence for $T_\parallel$ and $T_\perp$, respectively[22,23,27]; a linear bias dependence of the AP state resistance; and a constant P state resistance. We solve equation (1) for the instability of static OOP and IP states (see *Methods*). The solutions are plotted as solid lines, dashed lines and by the colour scale in Fig. 2 (c), respectively. We also use numerical integration to analyze ensuing dynamical states; the magnitude of the precession motion is defined as the root mean square of the difference of the time-varying component of magnetization along the $x$-axis and its mean value, $(m_x(t) - \langle m_x \rangle)_{RMS}$. This magnitude is directly related to the experimentally measured output power, as the RF signal from STNOs is to given by the time varying projection of $\mathbf{m}$ along the magnetization of the reference layer[10,28], in this case fixed along the $x$-direction (see Fig. 1 (a)). The average $m_x$ component of magnetization, obtained numerically, is plotted in Fig. 2 (d), along with the analytical critical lines for the static IP states. $m_x$ = -1 corresponds to the AP in-plane state and $m_x$ = 0 corresponds to either alignment of $\mathbf{m}$ along the $z$-axis or circular precession around the $z$-axis. As in the experiment, stable dynamics occur only when electrons flow from the free to the reference layer. For currents up to around 2.2 mA, the critical current for dynamics scales quasi-parabolically with external field (Fig. 2 (c)). However, above this value the region of precession turns back in on itself and dynamics are gradually quenched. Complete suppression of dynamics is achieved for currents above 3.5 mA. To be sure of the underlying cause, we set $\frac{\partial R_{AP}}{\partial V}$ = 0 $\frac{\Omega}{V}$ and this reproduces what is expected for metallic systems and no curvature is observed[3,10] (see Supplementary Information: Part 3). While a strict circular cross section was assumed for the analytical calculation, numerical simulations show that the curvature of the critical lines is not affected by varying the shape of the free layer.

The inclusion of the TMR bias dependence $\frac{\partial R_{AP}}{\partial V}$ is vital in order to explain the curvature of the critical lines in Fig. 2 (a). Comparing the data shown in (a,b) and (c,d) in Fig. 2, the analytical model reproduces qualitatively the main features of the experimental diagram, while numerical data confirms the precession occurs in the areas bound by the critical lines. The results also show that the angular dependence of resistance in MTJs (equation (2)) helps to sustain precession in the absence of any intrinsic asymmetry.

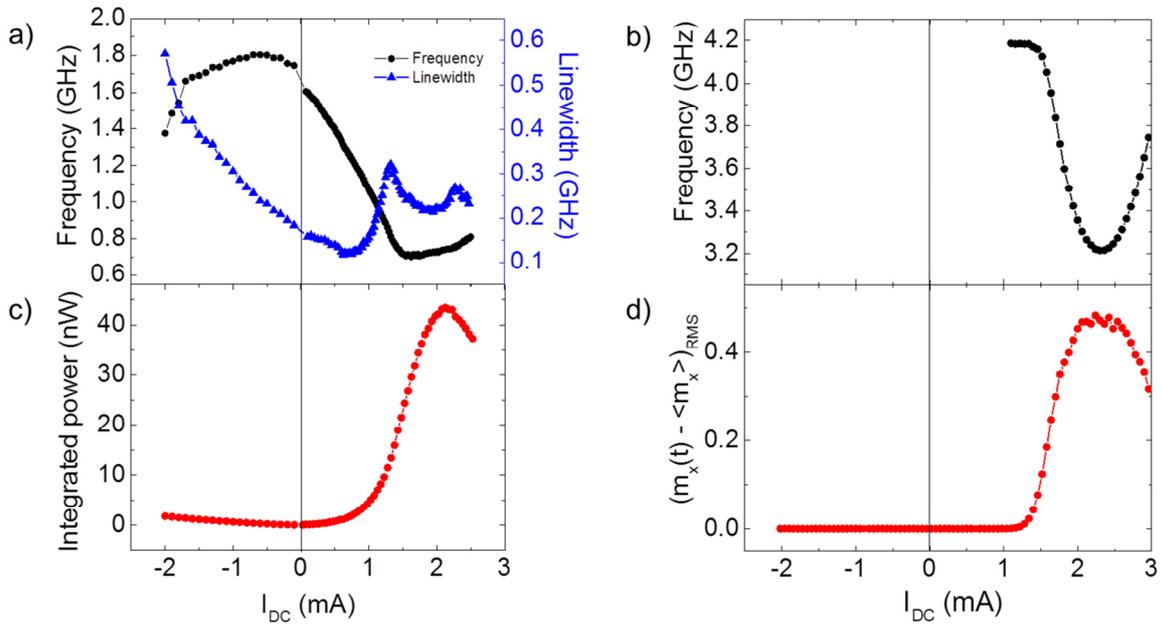

**Figure 3. Frequency characteristics of the STNO device.** Experimentally and numerically determined frequency and output power of the STNO device at $\mu_0 H_{ext}$ = 30 mT. Experimentally measured frequency and linewidth (**a**), and integrated power (**c**) of the STNO as function of $I_{DC}$. Numerically determined frequency (**b**), and dynamics intensity (**d**) as function of $I_{DC}$. In both cases, for positive current driving dynamics, we observe an initial rapid decrease in frequency, followed by a region of steady output power and frequency, and, finally a quenching of the output power. There is also an increase of the frequency due to the suppression of the spin-transfer torque asymmetry and ensuing quenching of the precession trajectory.

Further comparison can be made between experiment and numerical results by analyzing the expected precession frequency and output power. The measured frequency and output power as a function of $I_{DC}$ at $\mu_0 H_{ext}$ = 30 mT are shown in Fig. 3 (a) and 3 (c). Numerically obtained frequency and values of $(m_x(t) - \langle m_x \rangle)_{RMS}$ are shown in Fig. 3 (b) and 3(d).

According to Fig. 3 (a), for positive currents, the first signal is detected at 0.075 mA, at 1.6 GHz, corresponding to an estimated precession angle of 46° (see Supplementary Fig. S4). The frequency decreases with increasing current, reaching a minimum of 0.7 GHz at 1.5 mA, and then starts to increase again with a different slope. For positive current up to around 0.75 mA, the linewidth initially decreases with the current, which indicates that the increasing spin-transfer torque opposes the damping torque. For currents above 1 mA, the increase of the linewidth is a sign of increasing incoherency in the precession[29], as confirmed by the presence of the 1/f noise (as shown in Supplementary Fig. S5). The maximum linewidth is reached at 1.4 mA, where the estimated precession angle becomes larger than 90° (see Supplementary Fig. S4), indicating that, for currents above 1.4 mA, the macrospin approximation becomes invalid and inhomogeneous dynamics occur. This is further supported by the increase of the linewidth at higher currents, as well as the increasing 1/f tail (see Supplementary Fig. S5). It also has to be pointed out that the minimum frequency and maximum output power are reached at the same current in the simulation (2.25 mA). This is not the case in the experiment, further confirming the breakdown of the macrospin approximation under large applied currents.

For negative currents, a measured power is significantly lower (zero in the calculations). We attribute this to thermally-excited ferromagnetic resonance, also known as magnoise[13]. For this current polarity, the in-plane spin-torque acts as damping, yielding a significant increase of the signal linewidth with increasing current is observed. Note that this behaviour, as indeed the linewidth versus current dependence, is similar to the one exhibited by metallic samples[3].

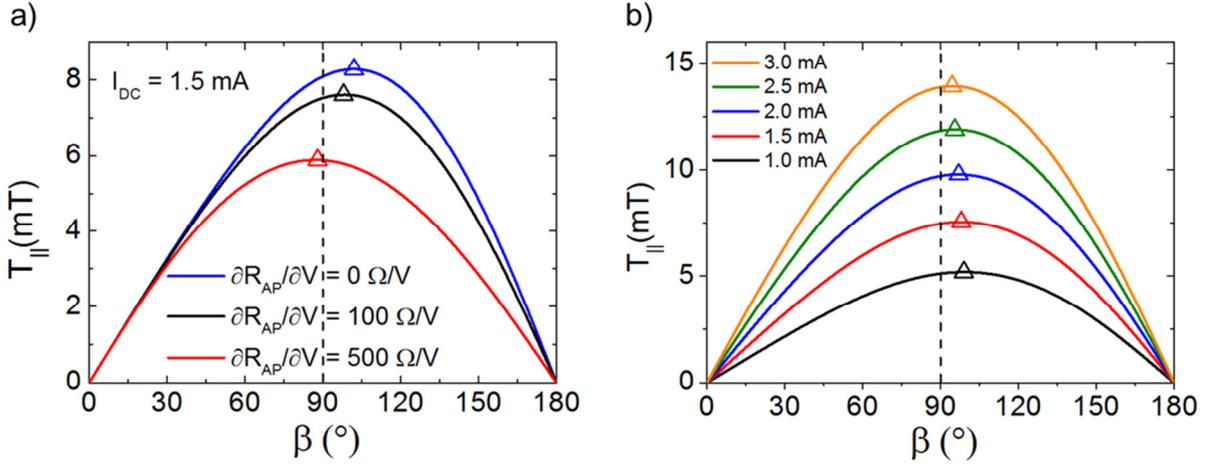

**Figure 4.** The in-plane component of the spin-transfer torque ($T_\parallel$) as function of the angle $\beta$ between the free and the reference layers in the STNO (as obtained using equation (3)). (**a**), The angular dependence of $T_\parallel$ for different values of $\frac{\partial R_{AP}}{\partial V}$ at $I_{DC}$ = 1.5 mA. An increase of $\frac{\partial R_{AP}}{\partial V}$ results in a reduction of the $T_\parallel$ angular dependence asymmetry. (**b**), The angular dependence of $T_\parallel$ for different values of DC current $I_{DC}$, and $\frac{\partial R_{AP}}{\partial V}$ = 100 $\frac{\Omega}{V}$. Increasing the applied current results in an increase of $T_\parallel$, but simultaneously a reduction of the $T_\parallel(\beta)$ asymmetry. The maximum of each curve is marked with an open triangle and $\beta$ = 90° is marked with a dashed line.

The analytical model also enables us to determine the dependence of $T_\parallel^{MTJ}$ on the angle between **m** and **p**, $\beta$, as well as the previously defined device parameters:

$$T_\parallel^{MTJ}(\beta) = \frac{\partial \tau_\parallel}{\partial V} \frac{R_P + \frac{\Delta R_0}{2}(1-\cos\beta)}{1+\frac{1}{2}|I_{DC}|\frac{\partial R_{AP}}{\partial V}(1-\cos\beta)} \sin\beta \cdot I_{DC}. \qquad (3)$$

Fig. 4 shows the asymmetry of the angular dependence of $T_\parallel$ (equation (3)) for different values of $\frac{\partial R_{AP}}{\partial V}$ at an applied DC current of 1.5 mA (Fig. 4 (a)), and for different values of $I_{DC}$ for $\frac{\partial R_{AP}}{\partial V}$ of 0, 100 and 500 $\frac{\Omega}{V}$ (Fig. 4 (b)). The blue line for $\frac{\partial R_{AP}}{\partial V}$ = 0 $\frac{\Omega}{V}$ shows the intrinsic spin-transfer torque asymmetry, with the maximum torque at a relative angle of 102°, arising solely from the cosine dependence of the resistance when experiments are conducted at a constant applied current and the in-plane spin-transfer torque scales as the corresponding voltage. Increasing the value of $\frac{\partial R_{AP}}{\partial V}$ to 100 $\frac{\Omega}{V}$ and 500 $\frac{\Omega}{V}$ shifts the maximum of $T_\parallel$ closer to 90° (98° and 88°, respectively). Indeed, increasing $\frac{\partial R_{AP}}{\partial V}$ reduces the TMR amplitude at the considered applied bias and hence counteracts the amplitude of the cosine oscillations of the resistance as function of angle, thereby decreasing the spin-torque asymmetry. For these parameters, we estimate that the asymmetry disappears for $\frac{\partial R_{AP}}{\partial V}$ = 330 $\frac{\Omega}{V}$, where the resistance of the antiparallel state becomes effectively equal to that of the parallel configuration.

Similar considerations explain the trends observed when analyzing the spin-torque angular dependence for fixed $\frac{\partial R_{AP}}{\partial V}$ (100 $\frac{\Omega}{V}$ in this case) and different values of the applied current (note a shift of the maximum from 99° to 94° corresponding to the current increase from 1 to 3 mA), see Fig. 4 (b). Indeed, one can observe an increase of the magnitude of $T_\parallel$ with the applied bias, since the magnitude of the torque is proportional to the current. In addition to that, the current is coupled with a reduction of the asymmetry in $T_\parallel(\beta)$, caused by the reduction of the TMR as the bias is increased. Consequently, an increase of the driving current of the STNO brings an enhancement of the output power up to certain current value as the precession angle is increased (reaching a maximum of 0.65 at 2.1 mA, see Fig. 3 (d)), above which the loss of asymmetry in $T_\parallel(\beta)$ and TMR become more relevant than the increase of the input power, leading to a decrease in precession angle and finally, the suppression of the dynamics at 3.25 mA.

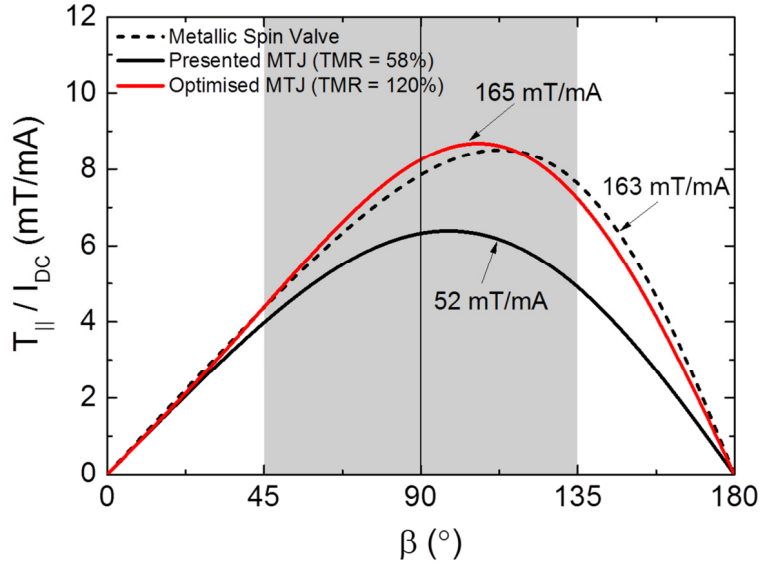

**Figure 5. Angular dependence of the spin-transfer torque efficiency $\frac{T_\parallel}{I_{DC}}$.** The in-plane spin-transfer torque efficiency ($\frac{T_\parallel}{I_{DC}}$) as a function of $\beta$, the angle between the free and the reference layer magnetizations. The black solid line is obtained using the parameters of the experimentally investigated MTJ. The red line corresponds to what is expected for a device with an increased TMR ratio of 120%. The dashed line represents the case of a metallic spin valve (using $T_\parallel^{GMR}$, as described in Ref.[3] with $\lambda$ = 1.5). The corresponding spin-torque efficiencies (expressed in mT/mA) are calculated using the *net* torque over one precession period. A circular trajectory is assumed with an angle of 45° about the *z*-axis. The resulting angle variation over one period is indicated with the shaded region.

Equation (3) is plotted as a function of $\beta$ in Fig. 5 for the experimental device parameters used in the analytical and numerical study (solid black line). The resulting torque displays an asymmetry around 90°, which allows for a *net* torque to be applied to **m** over one precession cycle. The degree of asymmetry is affected strongly by both $\frac{\partial R_{AP}}{\partial V}$ and $I_{DC}$ (see Supplementary Information: Part 5). Higher angular asymmetry of $T_\parallel^{MTJ}(\beta)$ results in a larger *net* torque, which competes with damping to sustain precession. Larger *net* torque yields a larger magnetization precession angle, which is directly proportional to the output power of the STNO device. With this in mind, we determine the *net* torque over one precession period for an angle of 45° about the *z*-axis (i.e., $45° < \beta < 135°$, the shaded region in Fig. 5). This is done by taking the difference of the integrated areas between 45° - 90° and 90° - 135°, and finally subtracting the damping torque. This then allows for the determination of the efficiency of STT-driven precession, expressed in units of mT/mA. For the experimentally investigated device this efficiency is 52 mT/mA. As a comparison to metallic systems, we also plot the angular dependence of $T_\parallel^{GMR}$ (using $\lambda$ = 1.5[3]) for the same magnetic free layer, which yields an efficiency of 163 mT/mA (black dashed line in Fig. 5). It can be seen that the efficiency of our MTJ device compares well to that of a metallic system, without the need for high applied currents or large out-of-plane magnetic fields. Equation (3) also allows us to investigate the effect of an increased TMR ratio on the angular dependence of $T_\parallel^{MTJ}$. The red line in Fig. 5 corresponds to a device with a TMR ratio of 120% (corresponding to $\Delta R$ = 228 Ω). This value is within the reported range for MgO-based MTJs[30], and results in an increased efficiency of 165 mT/mA.

**Conclusions**

We have shown that the bias and angular dependences of the magnetoresistance ratio in tunnel junctions plays a significant role in understanding spin-transfer driven magnetization dynamics. The angular dependence of the resistance difference between the parallel and antiparallel states introduces an angular asymmetry for the in-plane spin transfer torque parameter $T_\parallel^{MTJ}$ which helps to maintain steady-state precession. The bias dependence of the resistance, on the contrary reduces this asymmetry. These two mechanisms, in contrast to fully metallic systems, allow us to tune the asymmetry in $T_\parallel^{MTJ}$ as a function of current and to control the dynamical response of these devices. In comparison to metallic systems, tunnel junction systems have much larger power conversion ratios, $P_{out}/P_{in}$. For a typical metallic pillar $P_{out}$ is of order 1 nW, with typical driving currents of 10 mA. Tunnel junctions, on the other hand, have much larger $P_{out}$, up to 2 μW, with driving currents of about 1 mA, yielding a four order of magnitude increase in $P_{out}/P_{in}$ (see Supplementary Information: Part 5). The analytical model presented here qualitatively reproduces our experimental findings, including the curvature of the critical lines for dynamics, unique to magnetic tunnel junctions. It also allows us to make predictions for the device parameters which should be optimized in order to fast-track the exploitation of these devices as microwave transmitters and receivers. The bias dependence of the antiparallel state is not readily tuneable, however it has been shown to depend on the barrier properties[31]. A (symmetric or asymmetric) broken-linear TMR bias dependence can be caused by inelastic scattering, asymmetry of elastic tunnelling (reflecting the difference in a quality of the two barrier interfaces), first-order dependence of state density on energy, symmetry of distribution of inelastic tunnelling centers or a combination thereof. However, the assumption of a linear in-plane spin-torque bias dependence is only valid in the first two cases. Therefore, future research should aim at optimizing the barrier of MgO-based tunnel junctions not just for large TMR, but also to control the type of induced defects in order to limit the TMR bias dependence. Last but not least, taking the magnetoresistance bias dependence into account is going to become increasingly important. Indeed, the next generation of wireless devices does not only have to be tuneable, but should enable for the realization of technologies beyond 5G. While resonance frequencies of several hundred GHz can be achieved with MnGa-based Heusler half-metallic alloys[32,33], multilayers based on these compounds can exhibit a non-trivial TMR bias dependence such as non-monotonic or sign changing dependences[34].

**Methods**

**Analytical solutions for instability in the OOP and IP static states.** Equation (1) is solved for the instability condition of the static OOP state by using the trace and determinant of the Jacobian matrix $J$ (i.e., when $tr\,J > 0$ and $det\,J > 0$) at equilibrium positions in the small angle limit $\theta \to 0$ and $\theta \to \pi$. The critical lines are given by:

$$\mu_0 H_{ext}(I_{DC}) < \left| a_\parallel I_{DC} \frac{\Delta R_0 - |I_{DC}|\frac{\partial R_{AP}}{\partial V} R_P}{\alpha \left(2 + |I_{DC}|\frac{\partial R_{AP}}{\partial V}\right)^2} + a_\perp I_{DC}^2 \left( \frac{\Delta R_0(\Delta R_0 + 2R_P)}{\left(2 + |I_{DC}|\frac{\partial R_{AP}}{\partial V}\right)^2} - \frac{|I_{DC}|\frac{\partial R_{AP}}{\partial V}(\Delta R_0 + 2R_P)^2}{\left(2 + |I_{DC}|\frac{\partial R_{AP}}{\partial V}\right)^3} \right) - \mu_0 H_{k\perp} \right|. \quad (4)$$

The critical lines for the in-plane AP static state are defined by:

$$\mu_0 H_{ext}(I_{DC}) = \pm a_\parallel R_P I_{DC}. \quad (5)$$

The analytical and numerical solutions shown in Fig. 2 and Fig. 3 are calculated for the following set of parameters which is realistic in the case of our devices: $a_\parallel = 0.028\,\frac{T}{V}$, $a_\perp = 0.0008\,\frac{T}{V^2}$, $\alpha = 0.005$, $\Delta R_0 = 110\,\Omega$, $R_P = 190\,\Omega$, $\mu_0 H_{k\perp} = 120$ mT, and $\frac{\partial R_{AP}}{\partial V} = 105\,\frac{\Omega}{V}$.

**Numerical simulation.** Numerical integration of the LLGS equation (equation (1)) was performed using the MAPLE 8 program. We used the same parameters as in the analytical calculations (see above). The simulation enables for the evolution of the position of the magnetization vector under a defined set of parameters to be followed as a function of time. The initial magnetization direction of the free layer was set randomly, in order to take into account the bi-stability regions if they occur. The simulation time was 150 ns, and the final static or dynamic state was defined based on the last 2 ns of the simulation.

**Acknowledgments**

E.W., C.F., V.S. and A.M.D. acknowledge support from the Helmholtz Young Investigator Initiative Grant No. VH-N6-1048.

# *Supplementary Information*

## Tunnel magnetoresistance angular and bias dependence enabling tuneable wireless communication

Ewa Kowalska[1,2], Akio Fukushima[3], Volker Sluka[1], Ciarán Fowley[1], Attila Kákay[1],
Yuriy Aleksandrov[1,2], Jürgen Lindner[1], Jürgen Fassbender[1,2], Shinji Yuasa[3], Alina M. Deac[1]

[1] Helmholtz-Zentrum Dresden-Rossendorf, Institute of Ion Beam Physics and Materials Research, Bautzner Landstrasse 400, 01328 Dresden, Germany

[2] Institute of Solid State Physics, TU Dresden, Zellescher Weg 16, 01069 Dresden, Germany

[3] National Institute of Advanced Industrial Science and Technology, Spintronics Research Center, Umezono Tsukuba, 305-8568 Ibaraki, Japan

**Part 1: STNO output power**

The RF signal generated by an STNO device is detected by the spectrum analyzer as a voltage quantity per a defined frequency division. Thus, in order to express the spectral signal in the unit of power, we used the following formula:

$$PSD = \frac{V_{signal}^2 - V_{bg}^2}{Z \cdot RBW}. \tag{1}$$

Here, $PSD \left[\frac{W}{Hz}\right]$ is the Power Spectral Density, $V_{signal}$ is the voltage signal generated by the STNO sample, $V_{bg}$ is the background voltage, $Z$ is the impedance of the circuit (50 Ω), and $RBW$ is the Resolution Band Width (in this case, 3 MHz).

Supplementary Fig. S1 shows an example of data analysis for a single frequency spectrum of the STNO sample measured at +30 mT with a +2.3 mA current. The black line represents the measured spectrum, where the background signal has been previously subtracted. Since in this case a quite significant contribution from the low frequency noise is observed, the overall signal is fitted with Lorentz functions, so that two overlapping peaks can be distinguished: the main mode peak (red curve) and the low frequency noise (green curve). The total output power of the spin-torque oscillator is defined as the area under the red curve over the whole frequency. The power integration was conducted using a Matlab script based on the formula displayed in the inset of Supplementary Fig. S1.

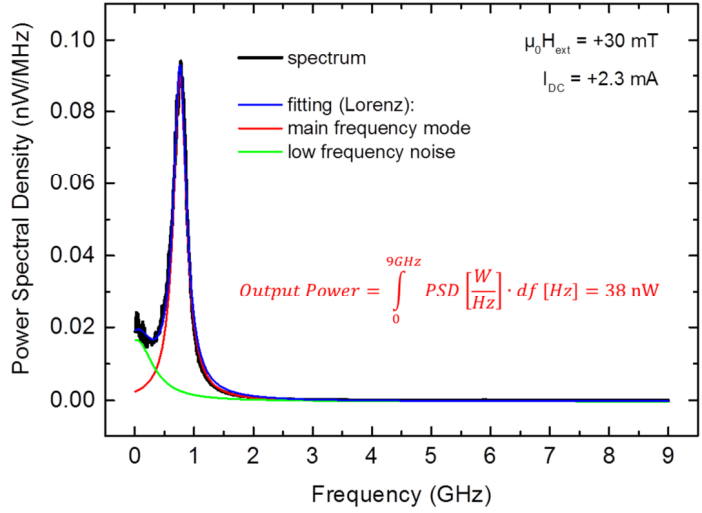

**Supplementary Figure S1.** An example of data analysis. The output power of spin-torque nano-oscillators is defined as an integral of the Power Spectral Density of the main frequency mode (red curve) over the frequency. Inset formula: $PSD$ - Power Spectral Density, $df$ - frequency step (in our case $df$ = 9 MHz).

**Part 2: Static resistance at dynamical states**

Simultaneously with the frequency spectra, we measured the static resistance versus field for every current value, shown in Supplementary Fig. S2. The magnitude of the TMR is directly proportional to the relative orientation of the magnetizations of the free and the reference layers, reaching its maximum for the AP state and minimum for the P state. Since the orientation of the reference layer magnetization is fixed, when dynamics are excited, the magnitude of the static resistance gives information about the $x$-component of the average position of the magnetization in the free layer (i.e., the position of the precession axis).

Supplementary Fig. S2 shows the evolution of magnetoresistance curves with the applied current. We observe here a reduction of the base resistance with increasing current, which results from the TMR bias dependence (see Fig. 1 (d) in the Main Text). For currents up to 0.8 mA, the TMR curve is similar to the curve measured at out-of-plane applied fields for a small probe current of 0.01 mA (see the red curve in Fig. 1 (c) in the Main Text). While approaching zero-field, we observe a dip in the resistance, which occurs due to the canting of the free layer magnetization, induced by an in-plane shape anisotropy of the nano-pillar (note that cross section of the nano-pillar is elliptical), combined with a slight parallel interlayer coupling with the reference layer. For currents above 0.8 mA, where the current density is high enough to drive magnetization dynamics, we observe an increase of the resistance close to zero-field, indicating a gradual tilting of the precession cone toward the in-plane AP orientation (i.e., $-x$ direction in Fig. 1 (a) in the Main Text).

Since the static resistance is proportional to the projection of the free layer magnetization on the magnetization vector in the reference layer, for every applied current, we calculate the change in resistance $\Delta R'|_{\mu_0 H_{ext}} = R(I_{DC}) - R(0.5\ mA)$ with respect to the resistance at the same field when using a small current $I_{DC}$ = 0.5 mA, which does not stimulate strong dynamics (see Fig. 3 (c) in the Main Text). The value of $\Delta R_0$ is directly proportional to $m_x$. The resistance change $\Delta R_0$ and its equivalent magnetic static state (expressed with averaged $m_x$ component of the magnetization in the free layer) are presented with the colour plot in Fig. 2 (b) in the Main Text. At small fields (0-30 mT), the average position of the magnetization tilts towards the antiparallel configuration (blue region); it finally reaches the AP state for currents above 2 mA (black region). This leads to the preliminary conclusion that the gap in the dynamics at small field, observed in diagrams in Fig. 2 (a), is actually an effect of the stabilization of the static in-plane AP state, under the influence of the spin-transfer torque, which favours the AP state for this current configuration.

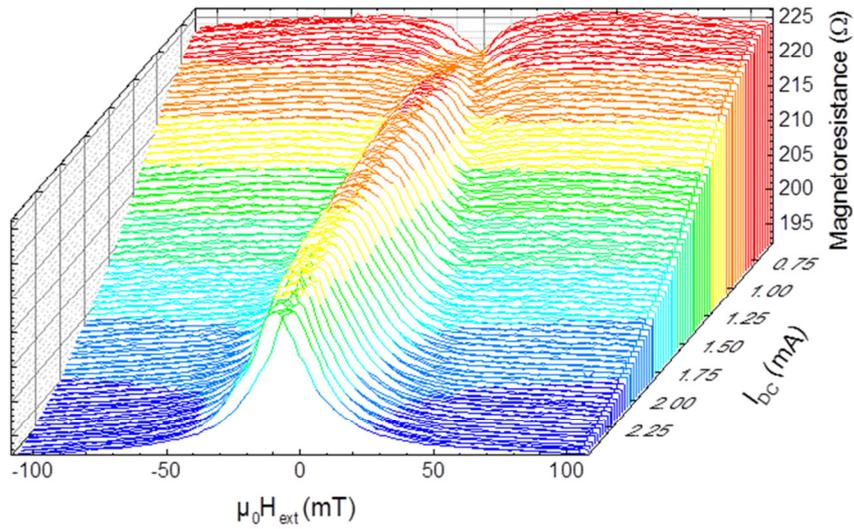

**Supplementary Figure S2.** Static magnetoresistance versus field for different current values. We observed a general decrease of the overall resistance with the current (due to the TMR bias dependence) and an increased resistance in the small field range (directly proportional to the magnitude of the applied current) indicating a gradual tilting of the precession cone toward the in-plane antiparallel direction.

## Part 3: Steady-state dynamics for $\frac{\partial R_{AP}}{\partial V}$ = 0 $\frac{\Omega}{V}$

In order to clarify the influence of the TMR bias dependence on the dynamical phase diagram of our STNOs, we initially performed analytical and numerical integrations of the Landau-Lifshitz-Gilbert-Slonczewski (LLGS) equation for the case of $\frac{\partial R_{AP}}{\partial V}$ = 0 $\frac{\Omega}{V}$ (i.e., no TMR bias dependence), for currents ranging from -3.5 to 3.5 mA (in 0.05 mA steps) and fields between -100 and 100 mT (with 2 mT increments). Supplementary Fig. S3 (a) displays the critical lines corresponding to the onset currents for precession (solid lines), as well as the quenching currents where the magnetization switches from a precessional state to an in-plane static state (dashed lines), as determined from analytical calculations (see *Methods* in the Main Text). The colour contrast marks the amplitude of the oscillations of the free layer magnetization along the direction defined by the magnetic moment of the reference layer, as obtained from numerical integrations of the LLGS equation. Consequently, the colour contrast should correlate directly with amplitude of the output signal that would be measured in an experiment conducted on such a device. The results are consistent with previous studies on metallic systems, where the intrinsic spin-torque angular dependence asymmetry was included[1]. Turning on the bias dependence of TMR, $\frac{\partial R_{AP}}{\partial V} > 0 \frac{\Omega}{V}$, the onset currents for precession exhibit a distinct curvature versus the applied field (see Supplementary Fig. S3 (b)), which has not been previously observed in metallic structures and thus constitutes a distinct signature for MTJs. Meanwhile, the quenching currents where the magnetization transits from steady-state precession to static in-plane states still depend linearly on the applied field (see dash-dot lines in Supplementary Fig. S3 (b)).

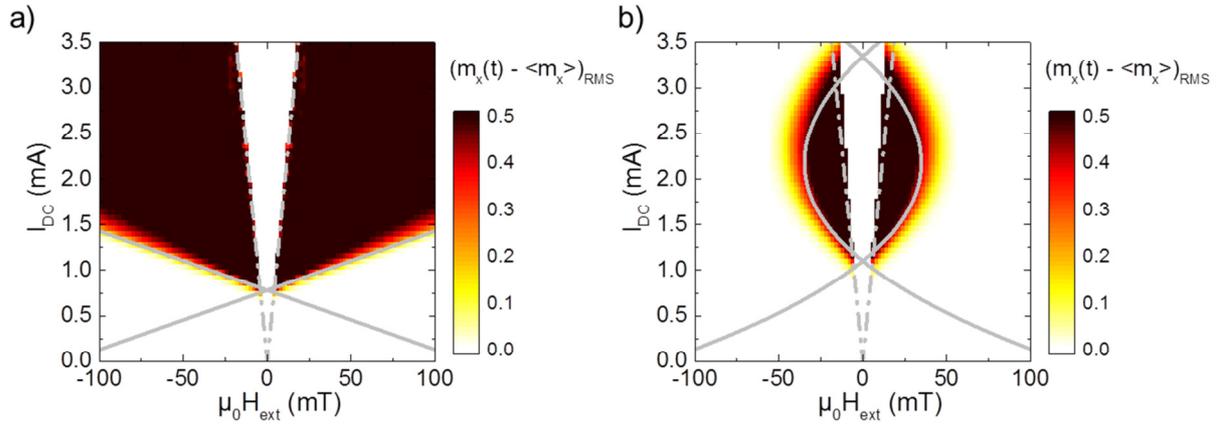

**Supplementary Figure S3.** Root mean square of the difference between the instantaneous unit magnetic moment along the $x$-axis (i.e., along the magnetic moment of the reference layer) and its mean value as a function of $I_{DC}$ and $\mu_0 H_{ext}$. (a) $\frac{\partial R_{AP}}{\partial V} = 0 \frac{\Omega}{V}$, and (b) $\frac{\partial R_{AP}}{\partial V} = 105 \frac{\Omega}{V}$. The plots are limited to positive currents, as no dynamics was observed for negative bias, although the full current range was considered for the calculations.

**Part 4: Linewidth and details of spectral features**

We estimate the magnetization precession angle $\theta$ from the experimentally obtained frequency as a function of $I_{DC}$, shown in Fig. 3 (a) in the Main Text, using the following formula:

$$\theta = arccos\left(\frac{\frac{2\pi f}{\gamma} - B_{ext}}{B_{k\perp}}\right). \quad (2)$$

Here, $f$ is the precession frequency, $\gamma = 1.76 \cdot 10^{11} \frac{rad}{s \cdot T}$ is the gyromagnetic ratio, $B_{ext}$ = 30 mT is the external field, and $B_{k\perp}$ = 120 mT is the effective out-of-plane anisotropy.

The derived angle is plotted in Supplementary Fig. S4. The increase of the angle above 90° indicates that the macrospin approximation breaks down above ~ 1 mA. A similar behavior was obtained in metallic systems[2]. This interpretation is further supported by the increase in 1/f noise (see Supplementary Fig. S5), occurring above this bias. 1/f noise is indicative that the dynamics are becoming increasingly less coherent[3].

It is also worth to note that equation (2) is only valid when assuming a constant precession angle $\theta$ for a given frequency value (the same approximation we also used in our analytical calculations, see *Methods* section in the Main Text). In the real system, as well as in the macrospin simulation, the precession angle can be assumed as constant only for low currents. For higher applied currents, the magnetization precession trajectory deviates from circular shape and becomes more quasi-elliptical.

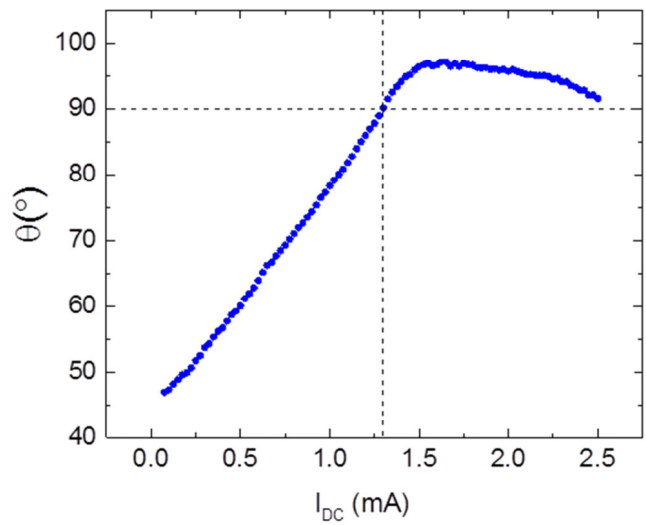

**Supplementary Figure S4.** Precession angle versus applied current for an applied field of 30 mT estimated using equation (2) from experimentally obtained frequency shown in Fig. 3 (a) in the Main Text.

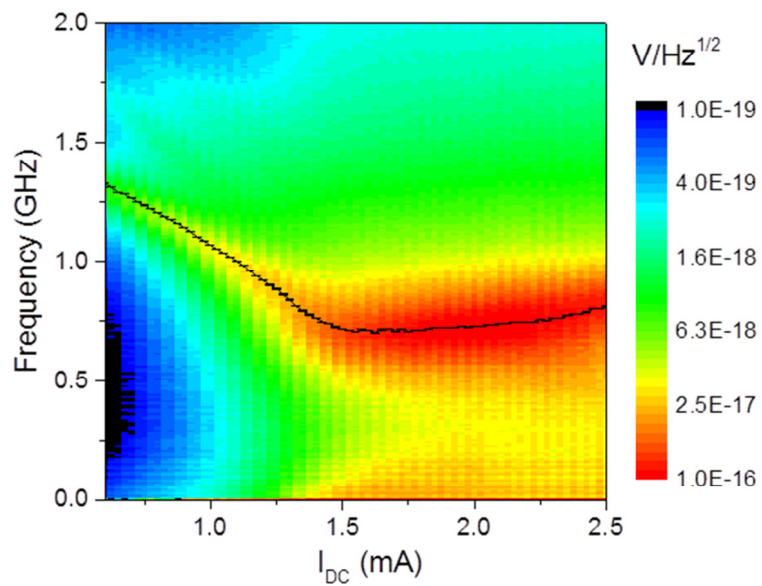

**Supplementary Figure S5.** Full spectra showing the increase in 1/f noise at the same applied current as the increase in linewidth in Fig. 3 (a) in the Main Text.

**Part 5: STNO efficiency: metallic spin valve versus magnetic tunnel junction**

The efficiency of STNOs can be expressed by the power conversion ratio of the output power (emitted by the device) to the input power (required to excite steady-state precession): $P_{out}/P_{in}$. Typically, for the same lateral size of the nano-pillar, the resistance of a metallic spin-valve is of the order of 10 Ω, and that of an Fe/MgO/Fe magnetic tunnel junction ~100 Ω. The characteristic operation DC currents are experimentally ~10 mA and ~1 mA, respectively. Thus we obtain following input power of these two types of STNOs:

$$P_{in}^{GMR} = RI_{DC}^2 = 10\ \Omega \cdot (0.01\ A)^2 = 10^{-3}\ W \tag{3}$$

$$P_{in}^{TMR} = RI_{DC}^2 = 100\ \Omega \cdot (0.001\ A)^2 = 10^{-4}\ W \tag{4}$$

According to the literature, the output power of STNOs is of order of nW for metallic devices[2], and µW for MgO-based systems[4,5]. Thus, the power conversion ratios are as follows:

$$\frac{P_{out}^{GMR}}{P_{in}^{GMR}} = \frac{10^{-9}\ W}{10^{-3}\ W} = 10^{-6} \tag{5}$$

$$\frac{P_{out}^{TMR}}{P_{in}^{TMR}} = \frac{10^{-6}\ W}{10^{-4}\ W} = 10^{-2} \tag{6}$$

Consequently, the power conversion efficiency of STNOs based on MgO-based magnetic tunnel junctions is four orders of magnitude larger than that of fully metallic devices.